\documentclass[submission, Phys]{SciPost}

\usepackage{graphicx,amssymb,soul,color,amsmath,bm}
\usepackage{epstopdf,hyperref}
\usepackage{braket,dsfont,nicefrac}
\usepackage[mathcal]{euscript}
\usepackage[normalem]{ulem}
\hypersetup{colorlinks=true,citecolor=blue,linkcolor=magenta}

\sethlcolor{yellow}
\allowdisplaybreaks

\usepackage{xcolor}

\newcommand{\rhohalf}{e^{-\frac{\beta}{2} H}}

\begin{document}

	\begin{center}{\Large \textbf{
				Enhanced Convergence of Quantum Typicality using a Randomized Low-Rank Approximation
	}}\end{center}
	
	\begin{center}
		Phillip Weinberg\textsuperscript{*}
	\end{center}
	
	\begin{center}
		Department of Physics, Northeastern University, \\
		Boston, Massachusetts 02115, USA
		\\
		* p.weinberg@northeastern.edu
	\end{center}
	
	\begin{center}
		\today
	\end{center}
	
	\section*{Abstract}
	{\bf 
		We present a method to reduce the variance of stochastic trace estimators used in quantum typicality (QT) methods via a randomized low-rank approximation of the finite-temperature density matrix $e^{-\beta H}$. The trace can be evaluated with higher accuracy in the low-rank subspace while using the QT estimator to approximate the trace in the complementary subspace. We present two variants of the trace estimator and demonstrate their efficacy using numerical experiments. The experiments show that the low-rank approximation outperforms the standard QT trace estimator for moderate- to low-temperature. We argue this is due to the low-rank approximation accurately represent the density matrix at low temperatures, allowing for accurate results for the trace.
	}

	\vspace{10pt}
	\noindent\rule{\textwidth}{1pt}
	\tableofcontents\thispagestyle{fancy}
	\noindent\rule{\textwidth}{1pt}
	\vspace{10pt}
	
	\section{Introduction}
	
    Quantum typicality (QT) methods are powerful tools used for studying finite-temperature physics with exact diagonalization (ED), however, they fall under a broad class of stochastic trace estimators having applications in many other fields~\cite{drabold93,jaklic94,aichhorn03,long03,weise06,avron11,sugiura13,hanebaum14,hyuga14,roosta-khorasani15,saibaba17,sugiura17,okamoto18,schnack20}. More sophisticated applications of QT range from dynamic quantum typicality (DQT) for calculating real-time dynamics at finite temperature~\cite{bartsch09,elsayed13}, to minimally entangled typical thermal states (METTS) used for calculating finite-temperature physics with matrix and tensor product states~\cite{white09,stoudenmire10,wietek19}. QT methods generally fall under two categories, each using a different approximation of the trace:
	\begin{gather}
		{\rm Tr}\left(O e^{\beta H}\right)\approx \frac{1}{M}\sum_{i=1}^M \langle z_i |e^{-\frac{\beta}{2} H} O e^{-\frac{\beta}{2} H} |z_i\rangle\label{eq:QT_sym}\\
		{\rm Tr}\left(O e^{\beta H}\right)\approx \frac{1}{M}\sum_{i=1}^M \langle z_i | O e^{-\beta H} |z_i\rangle.\label{eq:QT_asym}
	\end{gather}
	Here, $H$ is the Hamiltonian of the system, $\beta$ is the inverse temperature, and $|z_i\rangle$ are independent identically distributed random vectors in the relevant Hilbert space. To evaluate $e^{-\tau H} |z_i\rangle$, one employs Lanczos or some other method that can efficiently capture the action of the matrix exponential.
	
	The variance of the QT trace estimator is a major factor in the effectiveness of a given QT method. When the variance is high, more samples are required to obtain a certain precision for the trace~\cite{schnack20,sugiura13}. When applying QT with ED, the variance can depend very strongly on the temperature and even differ between Eqs.~\eqref{eq:QT_sym} and \eqref{eq:QT_asym}. At high temperatures, both Eqs.~\eqref{eq:QT_asym} and \eqref{eq:QT_sym} have a similar variance but Eq.~\eqref{eq:QT_sym} has a smaller variance as the temperature decreases~\cite{aichhorn03}. We will refer to Eq.~\eqref{eq:QT_sym} and Eq.~\eqref{eq:QT_sym} as low-temperature quantum typicality (LTQT) and high-temperature quantum typicality (HTQT) respectively. In the case of METTS, the variance is reduced by picking $|z_i\rangle$ as product states and using Markov chain Monte Carlo to sample product states with the largest weights~\cite{white09,stoudenmire10}. However, for ED, the Markov chain method for sampling $|z_i\rangle$ is too expensive due to the significant computational effort required to evolve a state in imaginary time. 
	
	While the random vectors act as a means to approximate the trace, one can also interpret QT as an approximation of the density matrix $\rho\propto e^{-\beta H}$. This observation becomes more apparent by inserting a complete set of states into Eq.~\eqref{eq:QT_sym} and rearranging the terms to form the trace over a density matrix defined as a statistical mixture of thermal pure states~\cite{hyuga14,sugiura17}. Some recent work proposed using a randomized low-rank approximation to reduce the variance of stochastic trace estimators~\cite{lin17,saibaba17,meyer20}. Randomized low-rank approximations are a class of algorithms that use random vectors and the action of the matrix-vector product operation to create a low-rank approximation of said matrix\cite{halko12}. 
	
	This paper will show that one can use thermal pure states to generate a randomized low-rank approximation of the canonical density matrix. We then show how to use the low-rank approximation to construct new trace estimators for QT and DQT. We will call the new variants: low-rank quantum typicality (LR-QT) and low-rank dynamic quantum typicality (LR-DQT). To demonstrate the advantage of LR-QT and LR-DQT, we use numerical experiments on the spin-$1/2$ XXZ chain. We show that our low-rank versions of QT perform as well as or better than standard QT methods using a similar amount of computational effort. We also show that our new estimators work exceptionally well at low to moderate temperatures where standard QT methods have difficulties, making our low-rank variants preferable to regular QT.
	
	The rest of the paper is organized as follows: In Sec.~\ref{sec:trace} discuss the low-rank approximation applied to the trace estimator. Then, in Sec.~\ref{sec:lrqt} we use those formal results to construct LR-QT and LR-DQT. In Sec.~\ref{sec:exp} we present the results of the numerical experiments on the spin-$1/2$ XXZ chain. Finally, in Sec.~\ref{sec:con}, we end the paper with a discussion of possible applications of LR-QT.
	
	\section{Low-Rank Trace Estimator}
	\label{sec:trace}
	
    In this section we briefly outline the algorithm presented in Ref.~\cite{meyer20}. First, we define a low-rank approximation of order $r$ for a $N\times N$ matrix $A$ as:  
	\begin{equation}
		A\approx Q\left[A\right]_r Q^\dagger,
	\end{equation} 
	where the columns of $Q$ are a set of $r$ orthonormal vectors and $[A]_r$ is the $r\times r$  matrix. There are many algorithms available to construct such a low-rank approximation; however, for this algorithm, the low-rank approximation is generated using a so-called randomized low-rank approximation. To construct $Q$ using a randomized low-rank approximation of $A$, one generates a set of $r$ independent identically distributed random vectors as columns in a $N\times r$ matrix $S$. Next, by applying the matrix $A$ on $S$, we obtain a new matrix $Y$. It has been proven that the span of the vectors in $Y$ closely approximates the rank $r$ subspace of $A$; therefore, one may obtain $Q$ by generating an orthonormal basis from the vectors stored in $Y$ using any number of methods~\cite{halko12}.  
	
	The trace of $A$ can be estimated using this low-rank approximation by breaking the calculation into two parts: the trace over the low-rank subspace spanned by $Q$ and the trace over the subspace that is complimentary to the span of $Q$. In the limit where the low-rank approximation becomes exact, the contribution from the complementary subspace will be small~\cite{meyer20}. As $r\ll N$, it is feasible to evaluate the trace of $A$ in the low-rank subspace exactly, but we can only estimate the trace over the complementary subspace. To accomplish this, we use the stochastic trace estimator by taking a set of $r$ identically distributed random vectors in the columns of the matrix $G$ and project the matrix onto the complementary subspace of $Q$ by evaluating:
	\begin{equation}
		\tilde{G} = G - QQ^\dagger G.\label{eq:G_proj}
	\end{equation}
	Using $\tilde{G}$ and $Q$ the trace can be estimated as follows~\cite{meyer20}:
	\begin{equation}
		{\rm Tr}(A) \approx {\rm Tr}\left(Q^\dagger A Q\right) + \frac{1}{r}{\rm Tr}\left(\tilde{G}^\dagger A\tilde{G}\right).\label{eq:lr_trace_A}
	\end{equation}
	The first term is the trace over the low-rank approximation of $A$ while the second term corresponds to the stochastic trace estimator applied to the projection of $A$ onto the complimentary subspace~\cite{meyer20}. The same analysis applies to matrix $A=B^2$, but we can modify the expression to make it symmetric if $B$ is Hermitian:
	\begin{equation}
		{\rm Tr}(B^2) \approx {\rm Tr}\left((BQ)^\dagger (BQ)\right) + \frac{1}{r}{\rm Tr}\left((B\tilde{G})^\dagger(B\tilde{G})\right).\label{eq:lr_trace_B}
	\end{equation}
	
	Some final notes on this algorithm related to generating the orthogonal basis $Q$ from $Y$. While Gram-Schmidt can be numerically unstable, it may be a good method for generating $Q$ because each column of $Y$ is independent. Thus, it is possible to iteratively generate the states in $Q$ until the error of the low-rank approximation falls below a predefined tolerance~\cite{halko12}. It is also possible to use a Cholesky decomposition of the overlap matrix $Y^\dagger Y$ to generate $R$ by noting that:
	\begin{equation}
	    Y^\dagger Y = (Q R)^\dagger (Q R) = R^\dagger R = LL^\dagger.
	\end{equation}
	Then, one can obtain $Q$ by inverting $R$ and solving $Y=QR$. While this approach might be appealing due to the significant decrease in the computational overhead in generating $R$, this method fails when the two or more vectors in $Y$ are linearly dependent. This is due to the Cholesky decomposition being ill-defined because $Y^\dagger Y$ is not positive definite. 
	
	\section{Low-rank Quantum Typicality}
	\label{sec:lrqt}
	
	In Eqs.~\eqref{eq:lr_trace_A} and \eqref{eq:lr_trace_B} if we replace $A\rightarrow e^{-\beta H}$ and $B\rightarrow e^{-\frac{\beta}{2} H}$ we two different approximations for the partition function. To calculate expectation values we could apply the trace estimator directly to $Oe^{-\beta H}$, or $e^{-\frac{\beta}{2} H}Oe^{-\frac{\beta}{2} H}$. However, we can approximate the trace of observables with the vectors used for estimating the partition function, similar to standard QT. Let us rewrite Eqs.~\eqref{eq:lr_trace_A} and \eqref{eq:lr_trace_B} in terms of the states stored in the columns of $\tilde{G}\rightarrow |\tilde{g}_i\rangle$ and $Q\rightarrow |q_i\rangle$. Let us also replace $A\rightarrow e^{-\beta H}$ and $B\rightarrow e^{-\frac{\beta}{2} H}$:
	\begin{gather}
	    {\rm Tr}\left(e^{-\beta H}\right)\approx \sum_{i=1}^r \langle q_i|e^{-\beta H}|q_i\rangle+\frac{1}{r}\sum_{i=1}^r\langle \tilde{g}_i|e^{-\beta H}|\tilde{g}_i\rangle\\
	    {\rm Tr}\left(\left(e^{-\frac{\beta}{2} H}\right)^2\right)\approx \sum_{i=1}^r \langle q_i|e^{-\frac{\beta}{2} H}e^{-\frac{\beta}{2} H}|q_i\rangle+\frac{1}{r}\sum_{i=1}^r\langle \tilde{g}_i|e^{-\frac{\beta}{2} H}e^{-\frac{\beta}{2} H}|\tilde{g}_i\rangle	    
	\end{gather}
	Now the similarities between the LR-QT and standard QT are more pronounced. The structure is identical except that in LR-QT, there are different types of vectors used in the two parts of the trace estimator. Drawing inspiration from regular QT, we can write down expressions for tracing over an operator written in terms of the original vectors used in calculating the partition function: 
	\begin{gather}
		{\rm Tr} \left(Oe^{-\beta H}\right) \approx  \sum_{i=1}^r \langle q_i|Oe^{-\beta H}|q_i\rangle+\frac{1}{r}\sum_{i=1}^r\langle \tilde{g}_i|Oe^{-\beta H}|\tilde{g}_i\rangle,\label{eq:lr_asym}\\
		{\rm Tr} \left(e^{-\frac{\beta}{2} H}Oe^{-\frac{\beta}{2} H}\right) \approx  \sum_{i=1}^r \langle q_i|e^{-\frac{\beta}{2} H}Oe^{-\frac{\beta}{2} H}|q_i\rangle+\frac{1}{r}\sum_{i=1}^r\langle \tilde{g}_i|e^{-\frac{\beta}{2} H}Oe^{-\frac{\beta}{2} H}|\tilde{g}_i\rangle.\label{eq:lr_sym}
	\end{gather}	
    Based on their structure, we will refer to Eqs.~\eqref{eq:lr_sym} and \eqref{eq:lr_asym} as low-rank low-temperature quantum typicality (LR-LTQT) and low-rank high-temperature quantum typicality (LR-HTQT), respectively. To extend the equations to study real-time dynamics simply replace $O\rightarrow O(t)$ and evolve the vectors $|q_i\rangle$ and $|\tilde{g}_i\rangle$ in the same way as one would do for DQT.

    Both LR-QT and LR-DQT can be summarized in the following steps:
	\begin{enumerate}
		\item[$(1)$] Generate random set of $r$ column vectors, $S$ and, calculate $Y = e^{-\beta H} S$ using Krylov, Chebyshev, etc.
		\item[$(2)$] orthogonalize $Y$ to express $Y = QR$.
		\item[$(3)$] Generate another set of $r$ random column vectors, $G$, and calculate $\tilde{G} = G - QQ^\dagger G$.
		\item[$(4)$] Calculate the partition function and ${\rm Tr} \left(Oe^{-\beta H}\right)$ using Eq.~\eqref{eq:lr_sym} or Eq.~\eqref{eq:lr_asym} by applying $ e^{-\beta H}$ or $\rhohalf$ to each vector in $Q$ and $\tilde{G}$, evolving in real-time for LR-DQT.
	\end{enumerate}
	The computational cost of LR-LTQT and LR-HTQT can be broken into three main pieces: $(i)$ calculation of the matrix exponential on a vector, $(ii)$ calculating the trace in Eq.~\eqref{eq:lr_sym} or Eq.~\eqref{eq:lr_asym}, and $(iii)$ performing QR decomposition and calculate $\tilde{G}$. steps $(i)$ and $(ii)$ are the same steps required in standard QT, while $(iii)$ is unique to LR-QT. At first glance, step $(iii)$ may prove to significantly increase the amount of effort required to calculate expectation values at multiple temperatures. For standard QT methods, one can obtain results for multiple temperatures using a single Lanczos basis per random vector with no extra effort~\cite{wietek19,krishnakumar19,schnack20}. For LR-QT, an intermediate step involves orthogonalizing the vectors in $Y$ to obtain $Q$, followed by the re-application of $e^{-\tau H}$ to $Q$. So the question becomes: is it possible to generate $e^{-\tau H}Q$ without explicitly applying the matrix exponential to the vectors in $Q$. As we will show, the answer is yes with a small amount of computational overhead. 
	
	First, let us assume that we have two families of vectors $Y(\tau)=e^{-\tau H} Y$ and $G(\tau)=e^{-\tau H} G$ with $Y$ and $G$ being two sets of $r$ independent identically distributed random vectors. Recall that for a given value of $\beta$, we must orthogonalize the vectors in $Y(\beta)$ in order to obtain $Q(\beta)$ leading to the decomposition:
	\begin{equation}
		Y(\beta) = Q(\beta) R(\beta).
	\end{equation} 
	Inverting this equation allows one to construct $Q(\beta)$ in terms of $Y(\beta)$. One might opt to use an RQ decomposition instead of a QR decomposition as it can be numerically more stable to invert the equation $Y^\dagger = R Q^\dagger$, regardless, the mechanics are the same for both methods. After evolving $Q(\beta)$ in imaginary time, we find:
	\begin{equation}
		e^{-\tau H}Q(\beta)=e^{-\tau H}Y(\beta)R(\beta)^{-1}=Y\left(\beta+\tau\right)R(\beta)^{-1}.
	\end{equation}
	To calculate the vectors required for the trace over the complimentary subspace we use Eq.~\eqref{eq:G_proj} to write:
	\begin{equation}
		e^{-\tau H}\tilde{G} = G\left(\tau\right) - \left(e^{-\tau H}Q(\beta)\right) Q(\beta)^\dagger G(0).
	\end{equation}
	Thus, we have overcome the issue of having to re-apply the matrix exponential for each temperature. The only extra cost here is the orthogonalization step for each temperature, as well as having to evolve states to $\tau_{\rm max}=1.5\beta_{\rm max}$ for LR-LTQT and $\tau_{\rm max}=2\beta_{\rm max}$ for LR-HTQT. If one uses a Cholesky decomposition on $Y^\dagger Y$ with the above procedure, it is possible to calculate LR-QT estimators without explicitly calculating $Q$ making LR-QT feasible for METTS as well as large scale ED calculations.

	\section{Numerical Experiments}
	\label{sec:exp}
	
    QT methods are powerful because Eq.~\eqref{eq:QT_asym} and Eq.~\eqref{eq:QT_sym} can give very accurate results with $M$ that is much smaller than the size of the Hilbert space. This behavior is due to the sample-to-sample variance of expectation values in thermal pure states decaying exponentially with increasing system size~\cite{sugiura13}. So far, we have only focused on the estimates for the trace; however, an expectation value requires taking the ratio of two trace estimates. To make our results practically relevant, we will focus on the variance of expectation values, averaged over independent realizations\footnote{By "variance" we are not discussing the variance of individual pure states, we are interested in the variance of the the estimate coming from a collection of pure states.}. To make a fair comparison between LR-QT and QT, we choose rank $r$ and $M=3r$ respectively so that the number of times we call the matrix exponential function is the same for each method. We focus primarily on the matrix exponential as that is the most expensive part of the calculation, making the two methods roughly equal in terms of computational complexity. 

    In our numerical experiments we will study behavior of both thermodynamic and time-dependent expectation values in the spin-1/2 $XXZ$ chain with periodic boundary conditions:
	\begin{equation}
		H = \sum_{i=1}^L (1+\Delta)S^z_i S^z_{i+1} + S^x_iS^x_{i+1}+S^y_iS^y_{i+1},\label{eq:H}
	\end{equation}
	where we have set the units such that $\hbar=k_B=1$. The observable we will measure is the nearest neighbor $S^z S^z$ correlator, 
	\begin{equation}
	    C = \frac{1}{L}\sum_{i=1}^L S^z_i S^z_{i+1}. \label{eq:nn_corr}
	\end{equation}
	For all cases considered, the trace is calculated over the sector with $\sum_{i}S^z_{i}=0$, and we fix the length of the chain to be $L=14$. We use full diagonalization to compute the matrix exponential, and we sample random vectors by drawing from the normal distribution for each entry.

	\begin{figure}[t]
		\centering
		\includegraphics[width=0.99\textwidth]{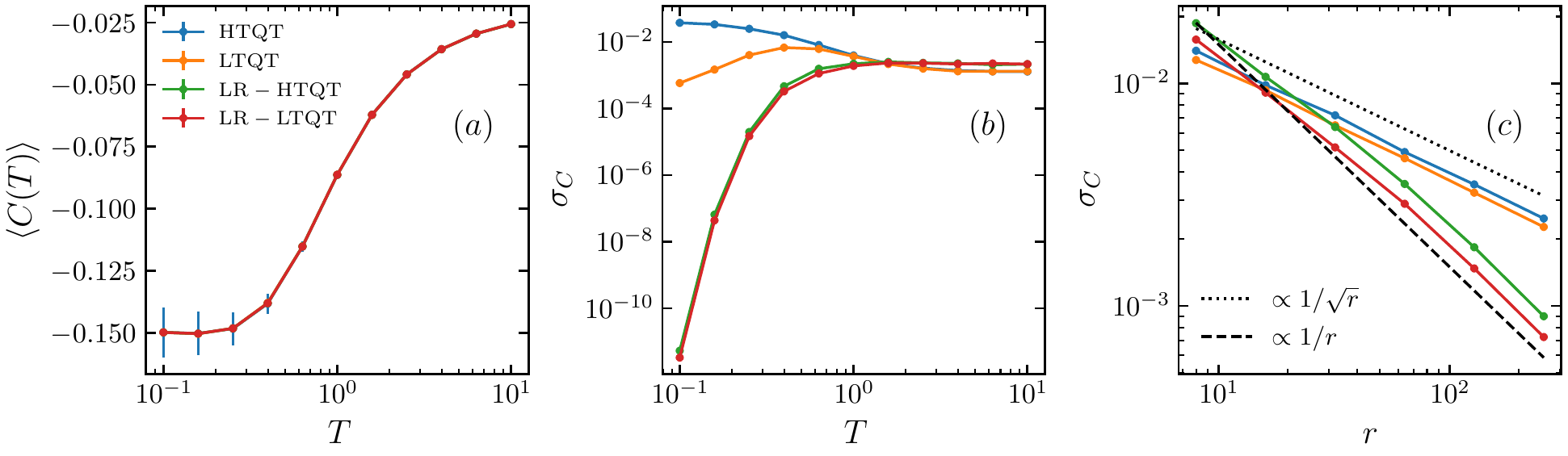}
		\caption{We show statistics of the trace estimator for all four QT methods applied to the nearest neighbor $S^z S^z$ correlator with $\Delta=0$. Panel $(a)$ shows the mean value of the four QT trace estimators with $r=10$, the error bars correspond to the variance. In panel $(b)$, we plot variance as function of temperature, $T$, with rank $r=100$, Finally, in  panel $(c)$: The variance as a function of $r$ with $T=1$. The dotted and dashed lines in panel $(c)$ correspond power-laws proportional to $1/\sqrt{r}$ and $1/r$ respectively. These lines represent the predictions for the scaling of the variance based on probabilistic error bounds cite and derived respectively in Ref.~\cite{meyer20}. All calculations shown are averaged over $1000$ independent realizations.} \label{fig:static}
	\end{figure}

	To begin, we show in Fig.~\ref{fig:static}$(a)$: the finite-temperature expectation value of Eq.~\eqref{eq:nn_corr} for $\Delta=0$ using both QT and LR-QT with $r=10$ as a function of temperature, $T$. The error bars correspond to the variance of the trace estimator calculated from $1000$ independent realizations of each method. All four methods give the exact result as we average over independent random realizations. As expected,  HTQT has the largest variance at low-temperatures while the error-bars are smaller than the markers for LTQT and the LR-QT methods. Next, we direct our focus to the variance of the trace estimators as a function of $T$ and $r$. In Fig.~\ref{fig:static}$(b)$ we show results as a function of $T$ for $r=100$ and $\Delta=0$. At high temperatures, the variance for the LR-QT methods is slightly larger than both LTQT and HTQT; however, as the temperature decreases, both LR-LTQT and LR-HTQT decrease monotonically. Compare this to the standard variants of QT, which have a greater variance as the temperature decreases.
	
	Another question is: how does the variance of LR-QT depend on $r$ compared to standard QT. In Fig.~\ref{fig:static}$(c)$, we plot the variance as a function of $r$ with $T=1$, again with $\Delta=0$. We also plot two lines corresponding to the scaling laws based on the probabilistic error-bounds presented in Ref.~\cite{meyer20}. While these error bounds are derived assuming a different distribution for the random vectors and a different trace estimator\footnote{compare Eqs.~\eqref{eq:lr_asym} and \eqref{eq:lr_sym} to Eq.~\eqref{eq:lr_trace_A} which is the formal result in reference~\cite{meyer20}}, the numerical results follow the scaling remarkably well. If it is true that those theoretical bounds apply here, it will prove that LR-QT scales are better than standard QT. The numerical results are promising, however, a rigorous proof would be nice to have.

	\begin{figure}[t]
		\centering
		\includegraphics[width=0.99\textwidth]{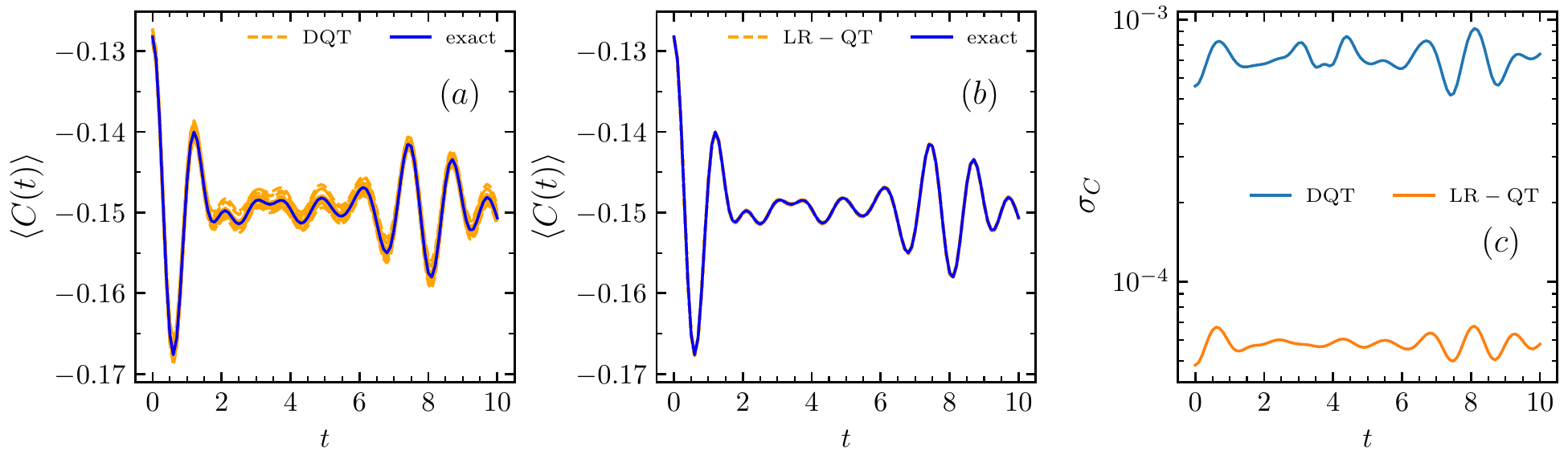}
		\caption{Here we show dynamics nearest-neighbor $S^z S^z$ correlator calculated as a function of time after a quench of the anisotropic parameter, $\Delta$, in the XXZ spin chain.  We prepare the initial state at $T=1/2$ with $\Delta=0$, and then the Hamiltonian is quenched to $\Delta=4$. We show results using full diagonalization, LR-QT, and DQT. In panel $(a)$ we plot the exact solution in blue along with 20 individual realizations calculated using DQT shown as dashed orange lines. Panel $(b)$ shows a similar plot to panel $(a)$ but for LR-LTQT instead. Panel $(c)$ shows the variance of the DQT and LR-QT methods as a function of time using $100$ independent realizations.}\label{fig:dyn}
	\end{figure}
	
   To benchmark LR-DQT, we study a protocol where we initialize the XXZ chain with $\Delta=0$ at $T=1/2$ and quench to $\Delta=4$. Same as before, we calculate expectation values of $C$, but now as a function of time after the quench. We present three different results from our calculations of both LR-DQT and DQT with $r=100$ in Figure.~\ref{fig:dyn}. Panel $(a)$ we plot the exact solution on top of $20$ realizations of DQT based on LTQT. Panel $(b)$ is similar to panel $(a)$ but plotting the $20$ realization of the LR-QDT based on LR-LTQT. Finally, in Figure.~\ref{fig:dyn}$(c)$ we show the variance of the trace estimator for $\langle C(t)\rangle$ as a function of time for both DQT and LR-DQT averaged over $100$ realizations. Just as we observed in the previous results for thermodynamic quantities, the LR-DQT method has a smaller variance than DQT. We specifically chose an example in which the variance was large enough to see by the eye. However, varying $T$ and $r$, we observed that LR-DQT is on par with or outperforms DQT. More importantly, when calculating real-time dynamics, the LR-DQT variant has the edge over standard DQT because one only has to evolve $2r$ vectors in real-time versus $3r$ vectors for DQT. Given that the most computationally expensive part of this calculation is the evolution in real-time, the $1/3$ reduction in cost is significant on top of the reduced variance.

	\section{Conclusion}
	\label{sec:con}

    In this paper, we have introduced LR-QT and LR-QDT to drastically improve the convergence of QT methods. The enhanced convergence is due to the usage of a randomized low-rank approximation of the density matrix. We have shown how to construct the low-rank approximation using existing thermal pure states from standard QT methods. Using numerical experiments on the spin-$1/2$ XXZ chain, we have shown that LR-QT outperforms standard QT when calculating thermodynamic quantities in low to moderate temperature regimes and we have shown that LR-DQT can give better results with less computational effort than standard DQT. We argue that the computational overhead required for LR-QT methods is small and generally worth the effort given the significant increase in the precision of expectation values.

	\begin{figure}[t]
		\centering
		\includegraphics[width=2.5in]{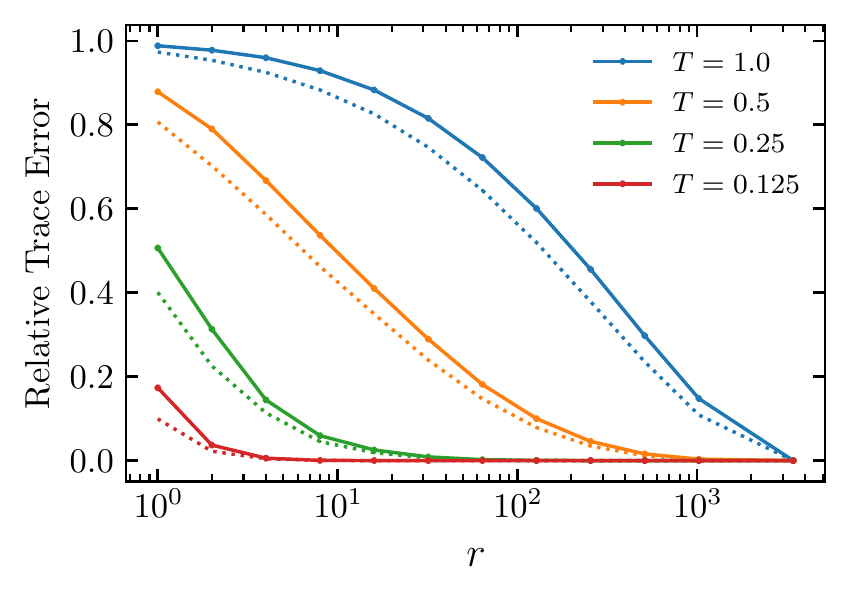}
		\caption{Relative errors for two different estimates of the partition function $Z={\rm Tr}\left(e^{-\beta H}\right)$ for various Temperatures. The solid lines corresponds to the trace of the low-rank approximation of $e^{-\beta H}$, the dotted line corresponds to the error in the exact trace of $e^{-\beta H}$ truncated at the $r$-th eigenvalue. The solid lines plotted here are averaged over $100$ independent realizations of the randomized low-rank approximation.}\label{fig:trace_err}
	\end{figure}

    An intuitive explanation for why LR-QT methods have a lower variance at low-temperatures is related to the spectrum of the density matrix, $e^{-\beta H}$. In this regime, the eigenvalues decay exponentially, allowing for a better low-rank approximation~\cite{halko12}. The increase in accuracy of the low-rank approximation implies that the contribution from the stochastic piece of the trace in Eq.~\eqref{eq:lr_asym} decreases. Therefore, any fluctuations that occur sample to sample are suppressed. To support this argument, we plot the relative between the trace of low-rank approximation of the density matrix and the exact trace as a function of rank $r$ and compare it to the relative error of the trace truncated the $r$-th eigenvalue. In Figure.~\ref{fig:trace_err} the solid lines correspond to the error of a single realization of the low-rank approximation while the dotted lines correspond to the truncated trace error. The plot clearly shows that the truncated trace and the low-rank approximation follow the same trend, supporting our argument. This argument is not the only reason for the success of LR-QT. While we do not have a formal proof, the numerical results in this work indicate that the LR-QT estimators follow the probabilistic error bounds derived for Eq.~\eqref{eq:lr_asym} in Ref.~\cite{meyer20}, indicating that LR-QT should, on average, scale better than QT as $r$ increases.

    Other than directly using LR-QT to study particular models, a natural application would be to combine LR-QT with numerical linked cluster expansions (NLCE)~\cite{rigol06,rigol07_1,rigol07_2,tang12}. In recent work, QT methods have successfully been applied to NLCE; however, one of the major obstacles was the high precision required to subtract out the contributions from all sub-clusters accurately. In Refs.~\cite{richter19,richter19_2} the calculation was possible in 1D because the series is particularly simply only having to subtract out one sub-cluster. In Ref.~\cite{krishnakumar19} the authors overcome the issue of precision by using Lanczos with full-orthogonalization. The Lanczos procedure generates a low-rank approximation using the Hamiltonian's extremal eigenvalues. The Hamiltonian's extremal eigenvalues can be mapped directly onto the density matrix's largest eigenvalues and provide an accurate approximation of the trace at low-temperatures, much like LR-QT. Finally, It is also worth noting that randomized low-rank approximations exist for a wide variety of matrix decompositions~\cite{halko12}. It may be helpful to apply these methods to other numerical methods beyond QT.

	\section{Acknowledgements}
	
	This work was supported by the U.S. Department of Energy (DOE), Office of Science, Basic Energy Sciences Grant Number DE-SC0019275. The authors would like to thank M. Bukov, A. Feiguin, and P. Patil for their useful comments and to D. Hendry for pointing out the relationship between the QR and Cholesky decompositions. All the calculations presented here where done using the QuSpin exact diagonalization package~\cite{weinberg17,weinberg19}. 
	
	\bibliographystyle{SciPost_bibstyle}
	\bibliography{refs.bib}
\end{document}